\documentclass[aps,prl,amsmath,amssymb,reprint,superscriptaddress]{revtex4-2}

\usepackage{graphicx}
\usepackage{xcolor}
\usepackage{mathtools}
\usepackage[normalem]{ulem}

\graphicspath{{./figures/}}

\newcommand{\la}{\langle}
\newcommand{\ra}{\rangle}

\DeclarePairedDelimiter{\norm}{\lVert}{\rVert}

\begin{document}

\title{Computing Quantum Mean Values in the Deep Chaotic Regime}

\newcommand{\majulab}{MajuLab, CNRS–UCA-SU-NUS-NTU International Joint Research Laboratory}
\newcommand{\cqt}{Centre for Quantum Technologies, National University of Singapore, 117543 Singapore, Singapore}
\newcommand{\LPTMS}{Université Paris-Saclay, CNRS, LPTMS, 91405, Orsay, France}
\newcommand{\PCS}{Center for Theoretical Physics of Complex Systems, Institute for Basic Science, Daejeon 34126, Korea}

\author{Gabriel M.~Lando}
\email{gmlando@ibs.re.kr}
\affiliation{\LPTMS}
\affiliation{\PCS}
\author{Olivier Giraud}
\affiliation{\LPTMS}
\affiliation{\majulab}
\affiliation{\cqt}
\author{Denis Ullmo}
\affiliation{\LPTMS}

\begin{abstract}
    We study the time evolution of mean values of quantum operators in a regime plagued by two difficulties: The smallness of $\hbar$ and the presence of strong and ubiquitous classical chaos.  
    While numerics become too computationally expensive for purely quantum calculations as $\hbar \to 0$, methods that take advantage of the smallness of $\hbar$ -- that is, semiclassical methods -- suffer from both conceptual and practical difficulties in the deep chaotic regime.
    We implement an approach which addresses these conceptual problems, leading to a deeper understanding of the origin of the interference contributions to the operator's mean value. We show that in the deep chaotic regime our approach is capable of unprecedented accuracy, while a standard semiclassical method (the Herman-Kluk propagator) produces only numerical noise. Our work paves the way to the development and employment of more efficient and accurate methods for quantum simulations of systems with strongly chaotic classical limits.
\end{abstract}


\maketitle

Semiclassical approximations to quantum mechanics, which we distinguish from purely ``classical approximations'' in the sense that they should account for interference effects, have a long and venerable history since they have been introduced almost immediately after the invention of quantum mechanics \cite{VanVleck1928} (and remarkably somewhat before \cite{einstein1917quantensatz}).  The successes of these semiclassical techniques in providing both effective approximation tools and physical understanding of the mechanisms at play have been demonstrated over time in a large number of systems in the integrable or nearly integrable regime \cite{Holthaus1994, *tannor2000semiclassical,*brodier2001resonance,  *tomsovic2018post, *ray2016dynamics, *lando2019quantum, *schlagheck2022resurgent} as well as in the deep chaotic regime \cite{tomsovic1991semiclassical, *tomsovic1993long, *sieber2001correlations, *heusler2007periodic}.

Surprisingly,  the semiclassical computation of some  very simple and common quantities poses both conceptual and practical problems that have not been solved until today. The conceptual problem has to do with the internal coherence of semiclassical approximations, namely the fact that different semiclassical approaches, or the use of different coordinate systems, must lead to the same result up to negligibly small terms. The derivation of this equivalence almost always implies the use of the stationary-phase (or steepest-descent) approximation (SPA), which appears therefore as one of the keystones on which we strongly rely to ensure the soundness of semiclassical approaches.

The problem is the following. For some physical quantities, among which the simplest is  the time evolution of a smooth operator's mean value, a blind application of the SPA leads to the incorrect result that only the classical contribution remains, and that all interference effects are washed out. This failure of the SPA, which goes beyond the need for uniform approximations \cite{berry1972semiclassical}, can actually be seen as one further example of the lack of commutation between the semiclassical ($\hbar \to 0$) and long-time ($t \to \infty$) limits \cite{voros1994aspects}. The practical consequence of this conceptual difficulty is that the computation of several quantities always relies on some form of initial or final value representation \cite{miller2001semiclassical, baranger2001semiclassical, *kay2005semiclassical, *ozorio2013initial}, the most famous one being Herman and Kluk's (HK) \cite{herman1984semiclasical}, which amounts essentially to evaluating all integrals using a numerical Monte Carlo technique. 
This family of approaches, and in particular HK, are easy to employ in practice and usually provide good approximations to correlations and spectra in systems whose classical dynamics is relatively close to integrability.  However, the computation of mean values using such methods converges slowly and has larger errors when compared to correlation functions \cite{lasser2017discretising}.  This is further worsened by the fact that, due to an inherent prefactor inversion, initial value representations fail as one enters the deep chaotic regime \cite{miller2001semiclassical, liberto2016importance}.  The main consequence of such problems is that there is nowadays no semiclassical tool capable of computing something as simple as  the time evolution of a smooth operator's mean value in the deep chaotic regime -- even in the near-integrable regime, no efficient methods exist.
Note that a pure quantum numerical approach is also usually computationally expensive in the deep semiclassical regime (for which semiclassical expressions are most accurate), due to the need of very fine grids \cite{lasser2020computing}.

The conceptual difficulty mentioned above has been addressed in a recent paper by some of us \cite{mittal2020semiclassical}.  The source of failure of the SPA was identified, and a canonically invariant expression for the mean value of smooth operators was derived in which only the integrals for which the SPA cannot be used remain to be performed numerically. Moreover, these integrals should be sampled only on a classical scale, and are therefore much easier to evaluate than the highly oscillatory (on a quantum scale) integrals in HK \cite{SwartThesis}. 
However, \cite{mittal2020semiclassical} did not assess the numerical efficiency or semiclassical accuracy of the approach, nor did it compare it with other widely used methods such as HK.
The goal of this Letter is to illustrate the effectiveness of this approach on a simple system, while simultaneously providing a rather general conceptual background on why it works. In the regime where one can apply HK, we demonstrate that our approach does at least as well and at a significantly lower numerical cost.
More importantly,  our approach retains its remarkable accuracy as one goes deeper in the chaotic regime, while the HK results become essentially numerical noise.

\begin{figure}[t]
    \centering
    \includegraphics[width=\linewidth]{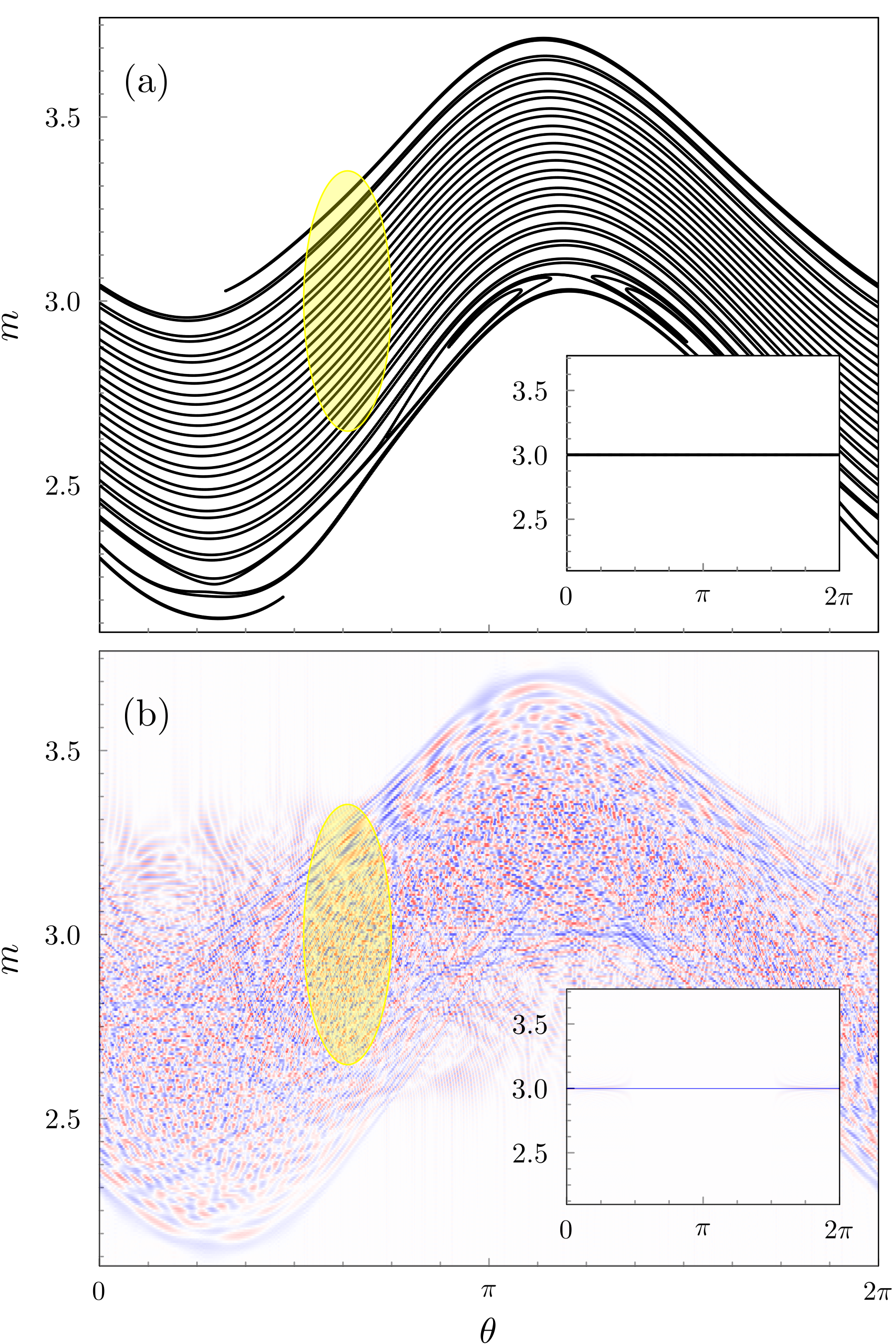}
    \caption{
    Comparison between (a) classical and (b) quantum evolutions obtained from iterating $\mathcal{L}_0$ and $\psi_0(\theta)$ using \eqref{eq:kicked_rotor_classical} and \eqref{eq:kicked_rotor_quantum}. For the quantum result we plot the Wigner function, whose positive and negative values are blue and red, respectively. Initial states $\mathcal{L}_0$ and $\psi_0(\theta)$ can be seen in the insets. The parameters chosen are $\alpha=0.6$, $M_0=3$ and $\hbar=M_0/500$. The yellow [deformed] circle marks 4 standard deviations of the classical detector \eqref{eq:W[O]}, which is centered at $(m_c,\theta_c)=(3, 2)$ and has $\sigma = 0.088$. For a zoom inside the detector, see Fig.~\ref{fig:wig_vs_lag_zoom}.
    }
    \label{fig:wig_vs_lag}
\end{figure}

As typical when employing semiclassical methods, our conclusions come accompanied by a wealth of geometrical insight into the underpinnings of mean-value calculations, tracing the source of quantum coherence to the self-interference of an evolving curve associated with the system's initial state. Here, we assume evolution to be generated by the kicked rotor system (KRS), 
\begin{equation}
\begin{cases}
    m_k = m_{k-1} - \alpha \, \cos \theta_{k-1} \\
    \,\, \theta_k = \theta_{k-1} + \alpha \, m_k \quad  \text{mod} \, 2\pi \,,
\end{cases} 
    \label{eq:kicked_rotor_classical}
\end{equation}
where $k$ counts the number of iterations, or \emph{kicks}, an initial point $(m_0, \theta_0)$ in phase space is subjected to \cite{chirikov1979universal}. The \emph{kicking strength} $\alpha$ is responsible for bringing the system from a near-integrable regime to a mixed one, and later to the deep chaotic regime in which the measure of regular islands tends to zero as phase space becomes almost fully chaotic. These regimes are reached for $\alpha \ll 1$, $\alpha \approx 1$ and $\alpha \gg 1$, respectively. Since the angular momentum $m$ is unbounded, a curve evolved according to \eqref{eq:kicked_rotor_classical} can diffuse in the deeply chaotic regime\sout{s}, while an initial state $\vert \psi_0 \rangle$ propagated using its exact quantum equivalent,
\begin{equation}
    \vert \psi_k \rangle =  \exp \left( -\frac{i \, \alpha \, \widehat{m}^2}{2\hbar} \right) \exp \left( -\frac{i \, \alpha \, \widehat{\sin \theta}}{\hbar} \right) \vert \psi_{k-1} \rangle \, ,
    \label{eq:kicked_rotor_quantum}
\end{equation}
will remain bounded \footnote{Note that while the angle operator is not well defined, its exponential (and, therefore, its sine and cosine) are perfectly rigorous, which is why we place the hat over the whole sine instead of only the angle. See \cite{levy1976afraid} for a nice exposition of the construction of angle and angular momentum operators.}. As is well-known, this renders the KRS a prototype for theoretical and experimental studies on dynamical and Anderson localizations \cite{lagendijk2009fifty, *floss2013anderson, *manai2015experimental}.
 
We will focus on the time-dependent mean value $\langle \widehat O \rangle(t=k) = \langle \psi_k \vert \widehat O \vert \psi_k \rangle$ with the initial state given by a plane wave of initial angular momentum $M_0$, 
\begin{equation}
\psi_0(\theta) = \frac{1}{\sqrt{2\pi}} \exp \left( - \frac{i \, M_0 \theta}{\hbar} \right) \, ,
\label{eq:psi0}
\end{equation}
with $M_0 = n \, \hbar$, $n \in \mathbb{N}$. The operator $\widehat O$ will be assumed to act as a ``classical detector'' that measures the probability of finding the system in the neighborhood of some angular momentum $m_c$ and some angle $\theta_c$, within some classical scale  $\sigma$. In practice $\widehat O$ is most easily defined through its Weyl symbol 
\begin{equation}\label{eq:W[O]}
        \mathcal{W} \big[ \widehat{O} \big](m, \theta) = \frac{1}{2\pi\sigma^2} \exp \left\{ - \frac{1}{2\sigma^2} \left[ (m - m_c)^2 + (\theta - \theta_c)^2 \right] \right\} \, ,
\end{equation}
which is just a symmetric, normalized phase-space Gaussian centered at $(m_c,\theta_c)$ and with standard deviation $\sigma$.
The larger we pick $\sigma$ the more classical the detector becomes, as can be seen by measuring the effective area enclosed by \eqref{eq:W[O]} in units of $\hbar$. We shall fix a small value for $\sigma$, namely $\sigma = 0.088$, in order to clearly observe a gradual improvement in the approximations as one moves deeper into the semiclassical regime.

Using \eqref{eq:W[O]} the quantum mean value of $\widehat{O}$ can then be calculated according to the phase-space average
\begin{equation}
    \la \widehat{O} \ra_\text{QM} = \int \text{d}m \, \text{d}\theta \, \mathcal{W}[\psi_k](m,\theta) \, \mathcal{W}\big[ \widehat{O} \big](m, \theta) \, ,
    \label{eq:mean_vals_QU}
\end{equation}
with 
\begin{equation}
    \mathcal{W} \left[ \psi_k \right](m, \theta) = \frac{1}{2\pi \hbar} \int_0^{2\pi} \!\!\!\! \text{d}\gamma \, \overline{\psi_k \left( \theta+\frac{\gamma}{2} \right)} \psi_k \left( \theta-\frac{\gamma}{2} \right) e^{\frac{i}{\hbar} \gamma m} 
\end{equation}
the Wigner function of $\psi_k$ (for the KRS the angular momentum is quantized, so that the integrals over $m$ should be understood as discrete sums \cite{supmat}). The semiclassical HK expression, $\la \widehat{O} \ra_\text{HK}$, is calculated analogously to $\la \widehat{O} \ra_\text{QM}$ but with the wavefunction $\psi_k$ replaced by its HK equivalent \cite{supmat}. 

The WKB quantization of the initial Lagrangian manifold $\mathcal{L}_0 = \{ (M_0, \theta_0), \, \theta_0 \in [0,2\pi)\}$ results exactly in expression \eqref{eq:psi0}. By iterating $\mathcal{L}_0$ classically using \eqref{eq:kicked_rotor_classical}, we produce an evolved Lagrangian manifold $\mathcal{L}_k = \{(m_k, \theta_k)\}$  that is the exact classical counterpart of \eqref{eq:kicked_rotor_quantum} with initial state \eqref{eq:psi0}. As an example, Fig.~\ref{fig:wig_vs_lag} displays a snapshot of the classical manifold $\mathcal{L}_{k}$ (top) and the quantum Wigner function $\mathcal{W}[\psi_{k}]$ (bottom) after $k=200$ iterations for the KRS in the near-integrable regime. The classical mean value, which we call $\la \widehat{O} \ra_\text{CL}$, is then given by substituting $\mathcal{W}[\psi_k]$ by a delta function with support on $\mathcal{L}_k$ in \eqref{eq:mean_vals_QU} and is known in physics literature as the truncated Wigner approximation (\emph{e.g.}~\cite{sinatra2002truncated, *polkovnikov2003quantum, *schlagheck2019enhancement}).

Following Maslov \cite{maslov2001semi}, our semiclassical mean value (SMV) approach is exclusively based on information contained in $\mathcal{L}_k$ and is given by
\begin{align}
    \la \widehat{O} \ra_\text{SMV} = \la \widehat{O} \ra_\text{CL}  &+ 2\sum_{\beta} \int \text{d} \eta \, \mathcal{W} \big[ \widehat{O} \big](\eta) a_+^\beta( \eta ) a_-^\beta( \eta ) \notag \\
    & \times \cos \left( \frac{i}{\hbar} S^\beta ( \eta ) + \frac{i \mu_\beta \pi}{2}  \right) \, ,
    \label{eq:SMV}
\end{align}
as adapted from \cite{mittal2020semiclassical}. Here, each $\beta$ entering the sum represents a pair of sections of $\mathcal{L}_k$ that fall inside the detector, which we denote by $\mathcal{L}_k^+$ and $\mathcal{L}_k^-$. We refer to each of these sections as a \emph{filament}. For each phase-space point $x_\pm = (m_\pm,\theta_\pm) \in \mathcal{L}_k^\pm$, the coordinates $\eta$ are formed as the centers $\eta = (x_+ + x_-)/2$. The center action $S^\beta$ is given by the symplectic area of the region enclosed by the path between $x_+$ and $x_-$ on $\mathcal{L}_k$ and the line segment joining $x_+$ and $x_-$ \cite{ozorio1998weyl, mittal2020semiclassical}. Along the path on $\mathcal{L}_k$, one is expected to cross several points at which the tangent space to $\mathcal{L}_k$ is parallel to the $m$ axis, which semiclassically gives rise to divergences in the $\theta$ representation, known as \emph{caustics} \cite{GutzwillerBook, *littlejohn1992van, *OzorioBook}. While \eqref{eq:SMV} is not impacted by them, it is fundamental to count how many were traversed in the form of a Morse index $\mu^\beta$ \cite{maslov2001semi, creagh1990geometrical}. Lastly, the coefficients $a_\pm$ measure how much the manifold was stretched during evolution, \emph{i.e.}~they can be thought of as the ratio between the length of an initial line segment in $\mathcal{L}_0$ and of its image in $\mathcal{L}_k$. For a detailed discussion of these concepts, see \cite{supmat}. 

The classical structure of Fig.~\ref{fig:wig_vs_lag}(a) is standard for near-integrable systems: The initial manifold $\mathcal{L}_0$ is stretched due to the shearing action of \eqref{eq:kicked_rotor_classical} when far from fixed points, but also captured near them, spiraling into whorls \cite{berry1979quantum, berry1979evolution}. The semiclassical interpretation of the oscillations seen in Fig.~\ref{fig:wig_vs_lag}(b) is that all the branches in $\mathcal{L}_k$ interfere with each other, forming the complicated patterns seen in the Wigner function. 
Before the so called \emph{Ehrenfest time} \cite{schubert2012wave}, classical and quantum evolutions produce the same expectation values (that is, quantum interference cancels out when computing mean values). Since we pick a large number of kicks, we are far beyond the Ehrenfest time \cite{schubert2012wave}, and the classical features of Fig.~\ref{fig:wig_vs_lag}(a) are blurred by quantum interferences in Fig.~\ref{fig:wig_vs_lag}(b).
By selecting a set of interfering branches in $\mathcal{L}_k$, detector \eqref{eq:W[O]} allows us to define a very objective notion of ``quantumness'': the mean value is all the more quantum as more interfering segments of $\mathcal{L}_k$ fall inside the detector.
The number of such filaments depends mostly on the curve defining the initial state and on the dynamics of the system, but for a high degree of chaoticity it will grow exponentially with $k$.  In Fig.~\ref{fig:wig_vs_lag}(a), for instance, there are 24 filaments within the yellow region defining 4 standard deviations of the detector.

\begin{figure}[t]
    \centering
    \includegraphics[width=\linewidth]{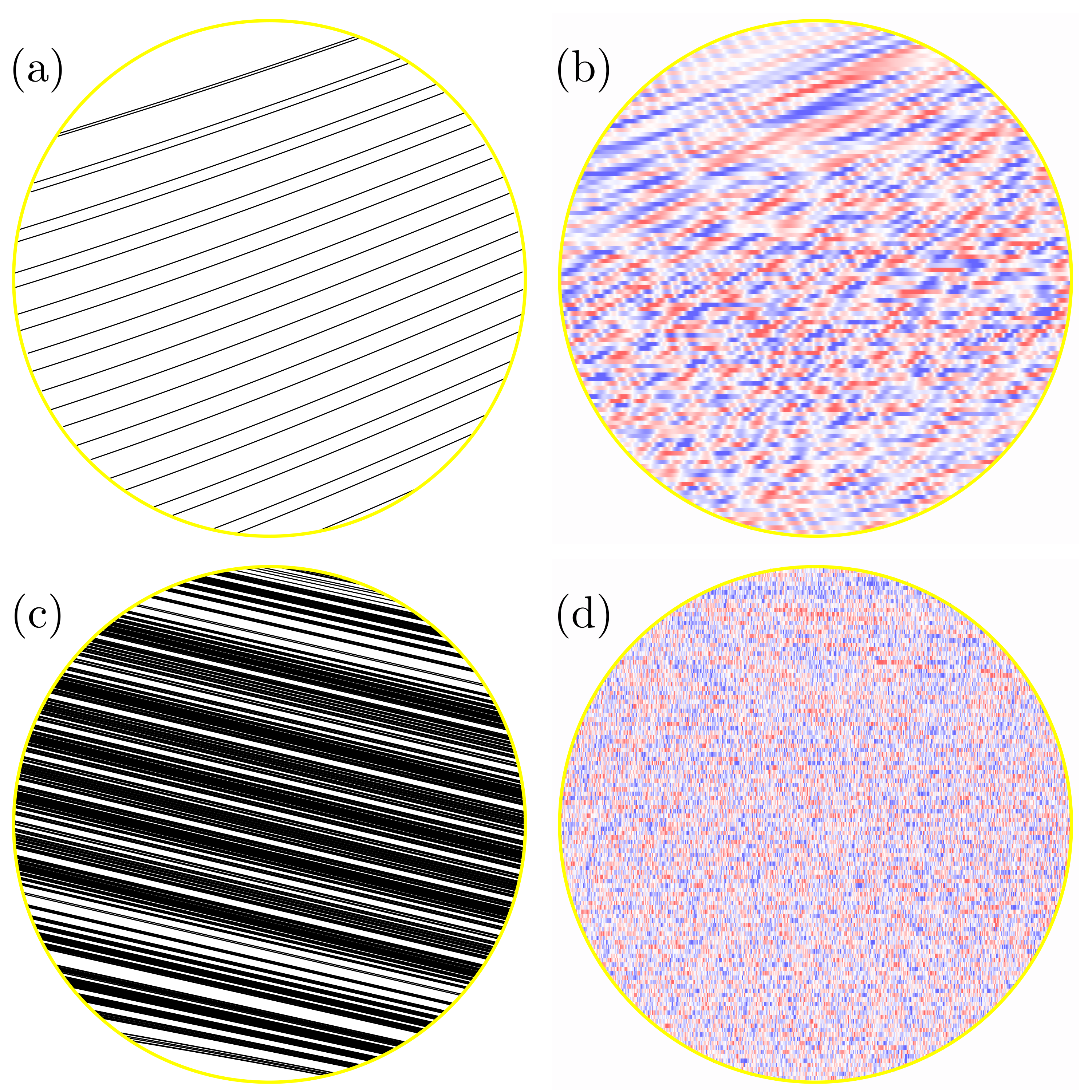}
    \caption{Panels (a) and (b) provide a zoom inside the detector of Fig.~\ref{fig:wig_vs_lag} for the classical and quantum cases, respectively. The semiclassical interpretation of the oscillations in (b) is that they are due to all 24 filaments in (a) interfering with each other. In (c) and (d) we provide an equivalent quantum-classical comparison in the deep chaotic regime instead of the near-integrable one, with $\alpha=4.0$ and $k=4$ ($\hbar$ and $M_0$ are unchanged). Panel (c) now consists of 851 filaments, whose interferences result in the ergodic-looking quantum structure displayed by the Wigner function in (d). Note that the discreteness of the Wigner functions is not due to pixelation, but a consequence of angular momentum quantization.}
    \label{fig:wig_vs_lag_zoom}
\end{figure}

\begin{figure*}[t]
    \centering
    \includegraphics[width=\linewidth]{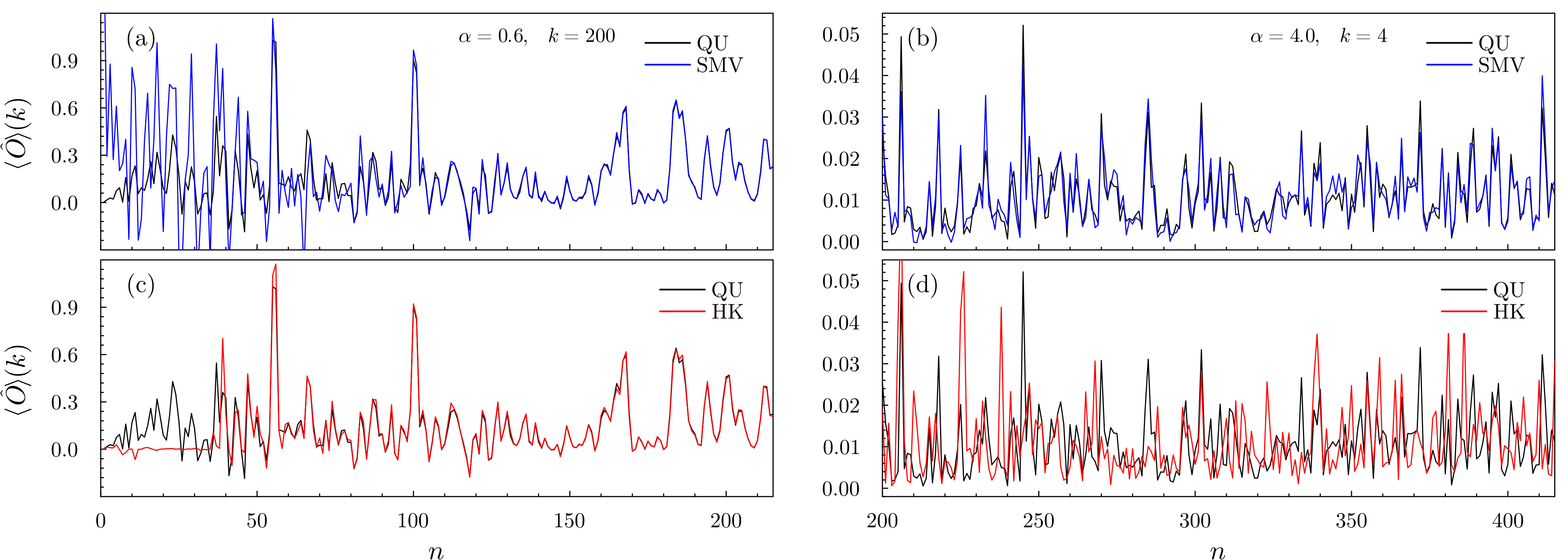}
    \caption{Mean value of operator \eqref{eq:W[O]} as a function of decreasing $\hbar$, with $\hbar = M_0/n$, computed quantum mechanically (QM), using the HK method, and our SMV formula \eqref{eq:SMV}. We connect the data with lines for visualization ease. Panels (a) and (c) are in the near-integrable regime, with $\alpha=0.6$ and $k=200$, and the detector defined as in Figs.~\ref{fig:wig_vs_lag} and \ref{fig:wig_vs_lag_zoom}. Panels (b) and (d) are in the deep-chaotic regime, with $\alpha=4.0$ and $k=4$. HK wavefunctions employed $5\times10^6$ and $4\times10^7$ trajectories for panels (b) and (d), respectively. Convergence tests and further comparisons can be found in \cite{supmat}.}
    \label{fig:hk_vs_smv.png}
\end{figure*}

The success of a semiclassical method will depend on how accurately it is able to reproduce pairwise interferences between filaments. This task becomes harder if the number of filaments increases, which can happen by increasing either the number of kicks, $k$, or the kicking strength, $\alpha$. From the detector's point of view, the long-time (large $k$) semiclassical dynamics and the deeply chaotic regime (large $\alpha$) are very similar, and the only important measure is the number of filaments falling inside its domain. Both limits, however, are well-known to be problematic for semiclassical mechanics \cite{voros1994aspects}, and producing reasonable quantitative results for a large number of filaments is a formidable task. To have a clearer idea on how intricate such a task is, in Fig.~\ref{fig:wig_vs_lag_zoom} we zoom in on the detector in order to have a better view of the classical and quantum structures inside it, both in the near-integrable regime of Fig.~\ref{fig:wig_vs_lag} and in the deep chaotic regime. The rather structured Wigner function Fig.~\ref{fig:wig_vs_lag_zoom}(b) in the near-integrable regime, semiclassically built on top of the simple filaments of Fig.~\ref{fig:wig_vs_lag_zoom}(a), finds an intimidating analogue in Fig.~\ref{fig:wig_vs_lag_zoom}(d). By emerging as a result of almost one thousand interfering filaments in Fig.~\ref{fig:wig_vs_lag_zoom}(c), the corresponding Wigner function looks ergodic and structureless. 

We expect a good match of both HK and SMV with the exact quantum mean values for the structure shown in Fig.~\ref{fig:wig_vs_lag_zoom}(a). This is indeed confirmed in Fig.~\ref{fig:hk_vs_smv.png}(a) and Fig.~\ref{fig:hk_vs_smv.png}(c), where we plot mean values as a function of $\hbar$. Here, since the accuracy of semiclassical approximations improves as $\exp(i \, M_0 \theta /\hbar )$ becomes highly oscillatory, instead of entering the semiclassical regime by increasing $M_0 = n \, \hbar$ we define a tunable $\hbar = M_0/n$, tracking our results as a function of the integer $n$. The advantage is that by fixing $M_0$ the classical dynamics is frozen.  We then see that in the near-integrable regime both HK and SMV become increasingly accurate as we decrease $\hbar$, with the former converging earlier but being slightly less accurate for smaller $\hbar$ values. On the other hand, in the deep-chaotic regime in Fig.~\ref{fig:hk_vs_smv.png}(d) the HK calculation is essentially numerical noise, while the SMV in Fig.~\ref{fig:hk_vs_smv.png}(b) displays a remarkable quantitative match with the quantum result. The result obtained from the HK method is also improved by the fact that we compute wavefunctions and only then the mean values. This allows us to renormalize the wavefunctions according to $\psi_\text{HK} \mapsto \psi_\text{HK} / \sqrt{\norm{\psi_\text{HK}}}$ before plugging them into \eqref{eq:mean_vals_QU}, correcting for normalization loss due to the fact that the HK propagator is not unitary. 
This procedure allows us to bypass one of the main flaws in the HK method, but can only be efficiently carried out for systems with a single degree of freedom.
It is also important to be sure that the HK wavefunctions are properly converged before taking the mean values, and we provide extensive convergence testing in \cite{supmat}. 

It can be argued that HK results could be improved by filtering trajectories or employing one of the many prefactor renormalization techniques developed  throughout the years in computational chemistry \cite{walton1996new, liberto2016importance}. These techniques, however, are only successful when there exist regular trajectories that survive the filtering process \cite{liberto2016importance}. In the deep chaotic regime phase space is completely dominated by the chaotic sea, and it is no surprise that even prefactor renormalization was not capable of improving the quality of HK calculations for the KRS -- even when used in conjunction with wavefunction renormalization \cite{Maitra2000}. The fact that the HK wavefunctions are converged, yet highly inaccurate, confirms that the complexity of the classical structure in Fig.~\ref{fig:wig_vs_lag_zoom}(c) is to blame: The HK method is simply incapable of accounting for all interfering filaments correctly. The large number of filaments, however, also leads to a combinatorial problem in a naive implementation of the SMV: Since all pairs enter \eqref{eq:SMV} and the number of pairs is given by $N(N-1)/2$, the computational cost increases quadratically with the number of filaments, $N$. This difficulty can be soothed by keeping in mind that not all pairs need to be included in \eqref{eq:SMV}, only the ones for which the action is not highly oscillatory -- in practice, of $\mathcal{O}(\hbar)$. This renders the computational cost linear in the number of filaments. Our implementation also relies on filaments not developing too much structure (such as coming out from the same direction as they came in, forming a \emph{finger}). This is by no means a shortcoming of the method itself,  and our algorithm can be improved and adapted to particular applications as our source code is made accessible to the public \cite{supmat}. For dealing with the previously mentioned fingers, for instance, a uniformization procedure in terms of Airy functions can be applied to \eqref{eq:SMV} \cite{berry1972semiclassical}.

In principle, our SMV method is ready to be generalized to the propagation of coherent states, where now complex  Lagrangian manifolds need to be employed \cite{tomsovic2015generalized, *tomsovic2016generalized}. 
An implementation for higher-dimensional systems is also possible, and will mainly imply that the intersection of the Lagrangian manifold defining our propagated state with the detector will be of dimension $d \ge 2$. Although these manifolds can be quite complicated, in the deep chaotic regime they are locally $d$-dimensional planes, just as the filaments of Fig.~\ref{fig:wig_vs_lag_zoom} are essentially lines. This can extend the applicability of our method to complex chaotic systems (\emph{e.g.}~two- or more electron atoms), which lie beyond the applicability horizons of initial value representations such as Herman and Kluk's.  Moreover our approach can be adapted to address the dynamics of hybrid quantum-classical systems \cite{singh2022hybrid}, or to compute other physical quantities with well-defined classical limits, such as the conductance matrix of quantum dots or the purity (linear entropy) of a bipartite quantum system.

\acknowledgements
We thank Alfredo Ozorio de Almeida and Steven Tomsovic for fruitful discussions. The authors gratefully acknowledge support from the Grant No.~ANR-17-CE300024. GML thanks Technische Universit{\"a}t Dresden for their hospitality and acknowledges financial support from the Institute for Basic Science in the Republic of Korea through the project IBS-R024-D1.

\bibliography{bib}

\end{document}

\onecolumngrid

\title{Supplemental material for ``Computing Quantum Mean Values in the Deep Chaotic Regime''}

\newcommand{\majulab}{MajuLab, CNRS–UCA-SU-NUS-NTU International Joint Research Laboratory}
\newcommand{\cqt}{Centre for Quantum Technologies, National University of Singapore, 117543 Singapore, Singapore}
\newcommand{\LPTMS}{Université Paris-Saclay, CNRS, LPTMS, 91405, Orsay, France}
\newcommand{\PCS}{Center for Theoretical Physics of Complex Systems, Institute for Basic Science, Daejeon 34126, Korea}

\author{Gabriel M.~Lando}
\email{gmlando@ibs.re.kr}
\affiliation{\LPTMS}
\affiliation{\PCS}
\author{Olivier Giraud}
\affiliation{\LPTMS}
\affiliation{\majulab}
\affiliation{\cqt}
\author{Denis Ullmo}
\affiliation{\LPTMS}

\maketitle

\section{Wigner-Weyl transforms on the cylinder }\label{sec:formulas}

We first recall the expressions of the Wigner-Weyl transforms in the  case of the usual phase space $(p,q) \in \mathbb{R}^2$. The Wigner function associated with a state $\ket{\psi}$ is given by
\begin{equation}\label{wpsi}
    \mathcal{W} \left[ \psi \right](p,q) =  \frac{1}{2\pi \hbar} \int_{-\infty}^\infty \!\!\!\! \text{d} x \, \overline{\psi \! \left( q+\frac{x}{2} \right)} \psi \! \left( q-\frac{x}{2} \right) e^{\frac{i}{\hbar} p x}\, ,
\end{equation}
and the  Weyl symbol associated with an operator is
\begin{equation}
    \mathcal{W} \big[ \widehat{O}\big](p,q ) =  \int_{-\infty}^\infty \!\!\!\! \text{d}x \, \bra{q -\frac{x}{2}}\widehat{O}\ket{q +\frac{x}{2}} e^{\frac{i}{\hbar} px}
\end{equation}
(note the different prefactor). We have the property that
\begin{equation}\label{trab}
    \tr(\widehat{A}\widehat{B})=\int_{-\infty}^{\infty}\frac{\textrm{d}p\, \textrm{d}q }{2\pi\hbar}
    \mathcal{W} \big[ \widehat{A}\big](p,q )\mathcal{W} \big[ \widehat{B}\big](p,q ).
\end{equation}
For $\widehat{A},\widehat{B}$  respectively given by an observable $\widehat{O}$ and a density operator  $\widehat{\rho}=\ketbra{\psi}{\psi}$, Eq.~\eqref{trab} becomes 
\begin{equation}\label{psiopsi}
   \tr(\widehat{O}\widehat{\rho}) = \bra{\psi}\widehat{O}\ket{\psi}=\int_{-\infty}^{\infty}\!\!\! \mathrm{d}p\, \mathrm{d}q  \,\,
    \mathcal{W} \big[ \widehat{O} \big](p,q ) \,\,\mathcal{W} \left[ \psi\right](p,q ) 
\end{equation}
(note again the different prefactor). 
 
In our case the quantum phase space is the cylinder $\{(m,\theta) \in \hbar \mathbb{Z} \times S^1 \}$ with $\theta$  an angle with periodicity $2\pi$ and the momentum discretized as $m=j\hbar$, $j\in\mathbb{Z}$. The integral in \eqref{trab} becomes $\int \mathrm{d}p \equiv \int \mathrm{d}m\to \hbar\sum_{m\in \hbar\mathbb{Z}}$. Then the Wigner transforms of a state and an operator respectively read
\begin{equation}
    \mathcal{W} \left[ \psi \right](m,\theta) =  \frac{1}{2\pi \hbar}\int_0^{2\pi}  \!\!\!\! \text{d}\gamma \, \overline{\psi \left( \theta+\frac{\gamma}{2} \right)} \psi \left( \theta-\frac{\gamma}{2} \right) e^{\frac{i}{\hbar} m\gamma}
 \end{equation}
and
\begin{equation}\label{weylop}
    \mathcal{W} \big[ \widehat{O}\big](m,\theta) = \int_0^{2\pi} \!\!\!\! \text{d}\gamma \, \bra{\theta-\frac{\gamma}{2}}\widehat{O}\ket{\theta+\frac{\gamma}{2}} e^{\frac{i}{\hbar} m\gamma}\, .
\end{equation}
Property \eqref{trab} becomes
\begin{equation}\label{trabdiscrete}
    \tr(\widehat{A}\widehat{B})=\frac{1}{2\pi}\sum_{m\in\hbar\mathbb{Z}}\int_0^{2\pi} \!\! d\theta
   \,\, \mathcal{W} \big[ \widehat{A}\big](m,\theta)\mathcal{W} \big[ \widehat{B}\big](m,\theta).
\end{equation}
     
\section{Quantum calculations}\label{sec:qu}

\subsection{Wavefunction evolution}

Quantum evolution is given by 
\begin{equation}
    \vert \psi_k \rangle = \exp \left( -\frac{i \, \alpha \, \widehat{m}^2}{2\hbar} \right) \exp \left( -\frac{i \, \alpha \, \widehat{\sin \theta}}{\hbar} \right)  \vert \psi_{k-1} \rangle \, . 
    \label{eq_app:kicked_rotor_quantum}
\end{equation}
In practice we employ a Fourier grid method, switching between angular momentum and angle representations by employing a fast Fourier transform, \textbf{FFT}, and its inverse, \textbf{iFFT}:
\begin{equation}
    \psi_{k}(\theta) = \text{\textbf{FFT}} \l\{ \exp \left( -\frac{i \, \alpha \, m^2}{2 \hbar} \right) \text{\textbf{iFFT}} \l[ \exp \left( -\frac{i \, \alpha \, \sin \theta}{\hbar} \right) \psi_{k-1}(\theta) \r] \r\} \, .
\end{equation}
Numerically, we compute the Fourier transform and its inverse by discretizing $\theta\in [0,2\pi[$  and restricting $m$ to the interval $[m_\text{min},m_\text{max}[$ with $m_\text{max} = -m_\text{min} = N\hbar/2$, namely
\begin{align}
        \theta &= s \, \Delta \theta \, ,  &\Delta \theta = \frac{2\pi}{N} \, , \quad s = 0, \, 1, \, \dots, N-1 \, . \label{eq_app:discrete_theta}\\
        m&=j \hbar,&j = -\frac{N}{2}, \, -\frac{(N-1)}{2}, \, \dots, 0, \, \dots, \frac{N-1}{2}.\label{eq_app:discrete_m}
\end{align}
 Taking $N=2^{10}$  for  $\alpha=0.6$ and $N=2^{13}$ for $\alpha=4.0$ ensured that for all values of $\hbar$ considered the whole wavefunction lies within the interval $[m_\text{min},m_\text{max}[$. 

\subsection{Computing mean values}\label{subsec:mean}

In theory, all mean values can be computed using the exact quantum formula expressed as an integral over phase space, with $\la \widehat{O} \ra_\text{QM}$ given by Eq.~\eqref{psiopsi}. Since a phase-space formulation is perfect for the ideas contained in our work, we adopted this view in the main text. In practice however, we need to use relatively large values of $N$ due to the smallness of our $\hbar$ parameter, which requires us to calculate $N^2$ entries for $\mathcal{W} \left[ \psi_k \right](m, \theta)$. It is therefore more computationally efficient to calculate mean values in the angle representation, \emph{i.e.~}
\begin{equation}
    \la \psi_k \vert \widehat{O} \vert \psi_k \ra = \int_{S^1 \times S^1} \text{d}\theta_1 \, \text{d}\theta_2 \, O_{\theta_1 \theta_2} \psi_k(\theta_1) \overline{\psi_k(\theta_2)} \, , \label{eq_app:means}
\end{equation}
where
\begin{equation}
    O_{\theta_1 \theta_2} = \la \theta_2 \vert \widehat{O} \vert \theta_1 \ra \, .
\end{equation}
Since the matrix $O_{\theta_1 \theta_2}$ also has $N^2$ entries, just as the Wigner function of $\psi_k(\theta)$, it might look like both strategies are equivalent. However, since the Wigner transform of the detector is localized around its center $(m_c, \theta_c)$, the matrix elements of $O_{\theta_1 \theta_2}$ will be non-negligible only in a small domain  of the $(\theta_1,\theta_2)$ space. A similar truncation procedure could also be performed with the Wigner functions, but truncating in the angle representation is significantly more straightforward. 

The first step to obtain the matrix element $O_{\theta_1 \theta_2}$ is to note that 
\begin{align}
    O_{\theta_1 \theta_2} = \la \theta_2 \vert \widehat{O} \vert \theta_1 \ra &= \int_{S^1} \!\!\! \text{d}\theta \, \la \theta_2 \vert \theta \ra \la \theta \vert \widehat{O} \vert \theta_1 \ra = \text{tr} \l( \widehat{P}_{\theta_2 \theta_1} \widehat{O} \r) \, ,
\end{align}
where we have defined  $\widehat{P}_{\theta_2 \theta_1} = \vert \theta_1 \ra \la \theta_2 \vert$. Using the Weyl representation, this gives
\begin{equation}
    O_{\theta_1 \theta_2} = \frac{1}{2\pi}\int_0^{2\pi}\!\!\! \text{d}\theta \sum_{m \in \hbar \mathbb{Z}} \mathcal{W}[ \widehat{P}_{\theta_2 \theta_1} ](m, \theta)\,\,\, \mathcal{W}[  \widehat{O} ](m, \theta) \, . \label{eq_app:symbols}
\end{equation}
The Weyl symbol of the detector is a Gaussian centered at $(m_c,\theta_c)$ and with standard deviation $\sigma$ (see main text). Using the definition \eqref{weylop} the Weyl symbol of $ \widehat{P}_{\theta_2 \theta_1}$ is
\begin{equation}
    \mathcal{W} \big[ \widehat{P}_{\theta_2 \theta_1}\big](m,\theta) =  \int_{0}^{2\pi} \!\!\!\! \text{d}\gamma \, \braket{\theta-\frac{\gamma}{2}}{\theta_1}\braket{\theta_2}{\theta+\frac{\gamma}{2}} e^{\frac{i}{\hbar} m\gamma}=e^{\frac{i}{\hbar} m(\theta_2-\theta_1)}\delta\left(\theta-\frac{\theta_1+\theta_2}{2}\right)\, 
\end{equation}
and \eqref{eq_app:symbols} results in
\begin{equation}
\label{o12finol}
    O_{\theta_1 \theta_2} = \frac{1}{(2\pi)^{3/2}\hbar\sigma}
    \exp\left\{ 
    -\frac{1}{2\sigma^2}  \left(\frac{\theta_1 + \theta_2}{2} - \theta_c\right)^2 
    - \frac{\sigma^2}{2 \hbar^2} (\theta_2 - \theta_1)^2
     + \frac{i}{\hbar} m_c (\theta_2 - \theta_1) \right\} \, .
\end{equation}
This expression is a Gaussian of width $\sigma$ along $(\theta_1 + \theta_2)/2$ and of width  $\hbar/\sigma$ along $(\theta_1-\theta_2)$, effectively truncating the   integrals in \eqref{eq_app:means}. 

\section{Semiclassical mean-value calculations}\label{sec:smv}

\subsection{Obtaining $\mathcal{L}_k$}\label{subsec:Lk}

As in all numerical calculations, our initial Lagrangian manifold $\mathcal{L}_0 = \{ (M_0, \theta_0) \, ;M_0 =\text{const.} \; , \theta_0 \in (0, 2\pi]\}$ needs to be discretized in order to be evolved into $\mathcal{L}_k = T^k \mathcal{L}_0  $, with the mapping $T$ defined as
\begin{equation}
T \begin{pmatrix}
    m \\ \theta
\end{pmatrix}
= \begin{pmatrix}
    m' \\ \theta'
\end{pmatrix} \qquad {\rm such \; that} \qquad
\begin{cases}
    m' = m - \alpha \, \cos \theta \\
    \,\, \theta' = \theta + \alpha \, m' \quad (\text{mod} \, 2\pi)
\end{cases} \; .
    \label{eq_app:kicked_rotor_classical}
\end{equation}
A numerical difficulty in obtaining the final manifold $\mathcal{L}_k$ is that different initial points will be propagated along trajectories that can have vastly different Lyapunov exponents. Therefore, a uniform discretization of $\mathcal{L}_0$, \emph{i.e.}~a uniform discretization of the initial variable $\theta_0$, will result in a highly non-uniform final manifold $\mathcal{L}_k$. In fact, different points in $\mathcal{L}_k$ might be so far apart that we entirely ``skip'' the detector (which we assume to have an effective radius of $4\sigma$) and miss a filament, leading to very large errors when computing mean values. 
To remedy this, we devise a predictor-corrector algorithm based on linearized dynamics that chooses adaptively the step size $\delta \theta$ on $\mathcal{L}_0$ so that final points on $\mathcal{L}_k$ are never more than $L_\text{max} (1 + \epsilon)$ apart from each other. Here, $L_\text{max}$ is the step size aimed on the final manifold (typically small compared with the size of the detector) and $\epsilon \ll 1$ is a security factor.

To proceed, let us assume the $j$ first points $\theta^{(0)}_0 \smeq 0,  \theta^{(1)}_0,\cdots,\theta^{(j-1)}_{0}$ of the discretization  of $\mathcal{L}_0$ have been determined, together with the corresponding final points $v^{(i)} = T^k (M_0 , \theta^{(i)}_{0})^t$  on $\mathcal{L}_k$, as well as the stability matrices $\mathcal{M}^{(i)}$ corresponding to the linearization of $T^k$ at $(M_0,\theta_0^{(i)})$, $0 \leq i<j$.   At linear order a point $(M_0, \theta^{(j-1)}_0+\delta\theta)$ will be mapped  to
\begin{equation}
    v^{(j)}_\text{predicted} = v^{(j-1)} + \delta \theta \, u^{(j-1)} \, \quad \text{where} \quad     u^{(j-1)}  =
    \mathcal{M}^{(j-1)}
    \begin{pmatrix}
    0 \\
    1
    \end{pmatrix} \; .
\end{equation}
Therefore, at linear order, choosing $\delta \theta = L_\text{max}/\left\vert u^{(j-1)} \right\vert$  guarantees that $\Delta \equiv \left\vert v^{(j)}_\text{predicted} - v^{(j-1)} \right\vert_{\mathbb{R} \times S^1}  =   L_\text{max}$. It therefore only remains  to check that  $\delta\theta$ is small enough that the error $ \left\vert v^{(j)}_\text{predicted} - v^{(j)} \right\vert_{\mathbb{R} \times S^1}$ between the true image $v^{(j)}$ of the point $(M_0 , \theta^{(i)}_{0})$ under map $T^k$ and its linear approximation $v^{(j)}_\text{predicted}$ only marginally changes that distance $\Delta$, \emph{i.e.} by a factor at most $(1+\epsilon)$.

Starting with $L = L_\text{max}$, the algorithm to determine $\theta^{(j)}_{0}$ is therefore as follows:
\begin{enumerate}
    \item Take  $\delta \theta = L/\left\vert u^{(j-1)} \right\vert$
    \item Compute  $v^{(j)}_\text{predicted} = v^{(j-1)} + \delta \theta \, u^{(j-1)}$  and $\tilde v^{(j)} = T^k (M_0,\tilde \theta^{(j)})$, with $\tilde \theta^{(j)} = \theta^{(j-1)} + \delta \theta$ the trial angle
    \item Then 
        \begin{itemize}
            \item If $ \left\vert v^{(j)}_\text{predicted} - \tilde v^{(j)} \right\vert_{\mathbb{R} \times S^1} < \epsilon \, L_\text{max}$:  choose  $\theta^{(j)} = \tilde \theta^{(j)}$,   $v^{(j)} =\tilde v^{(j)}$, compute  $\mathcal{M}^{(j)}$  and proceed to $\theta^{(j+1)}$ ;
            \item Else: start again from step 1 with $L=L/2$.
        \end{itemize}
\end{enumerate}
This algorithm allows us to obtain a discretized $\mathcal{L}_k$ whose points are at most $(1+\epsilon) L_\text{max}$ apart from each other, such that no filaments will be missed as long as $L_\text{max}$ is much smaller than the classical detector's radius. Since strong chaos means that $\mathcal{L}_k$ might have a length thousands of times that of $\mathcal{L}_0$, we also refrain from saving the full $\mathcal{L}_k$ manifold, and only keep the filaments falling inside the classical detector.

\subsection{Truncated Wigner approximation}\label{subsec:TWA}

The semiclassical mean value formula requires us to compute the constant classical contribution, $\la \widehat{O} \ra_\text{CL}$. This is done using the truncated Wigner approximation (TWA):
% this is correct! TWA knows nothing about momentum quantization
\begin{equation}
    \la \widehat{O} \ra_\text{CL} = \int_{S^1} \text{d}\theta_0 \int_\mathbb{R}\text{d}m_0 \, W_0(m_0, \theta_0) \, \mathcal{W}[\widehat{O}](T^k (m_0, \theta_0)) \, , \label{eq_app:preTWA}
\end{equation}
where $W_0$ is the Wigner function of the initial state, and $\mathcal{W}[\widehat{O}]$ is the Weyl symbol of $\widehat{O}$. Given that our initial state is just a delta in angular momentum space, its Wigner function is quite trivial:
\begin{equation}
    W_0(m_0, \theta_0) = \frac{1}{2\pi}  \delta(m_0 - M_0) \, ,
\end{equation}
such that \eqref{eq_app:preTWA} simplifies to 
\begin{equation}
    \la \widehat{O} \ra_\text{CL} =\frac{1}{2\pi} \int_{S^1} \text{d}\theta_0 \, \mathcal{W}[\widehat{O}](T^k (M_0, \theta_0))  \, . \label{eq_app:TWA}
\end{equation}
In practice, since $\mathcal{W}[\widehat{O}]$ is localized, we calculate it using only the filaments described in Sec.~\ref{subsec:Lk}.

\subsection{Implementing the final formula}\label{subsec:osc}

\begin{figure}
\begin{minipage}{.5\textwidth}
  \includegraphics[width=.7\linewidth]{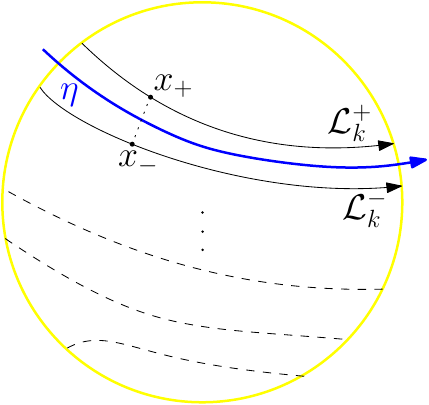}
\end{minipage}%
\hfill
\begin{minipage}{.5\textwidth}
  \includegraphics[width=.9\linewidth]{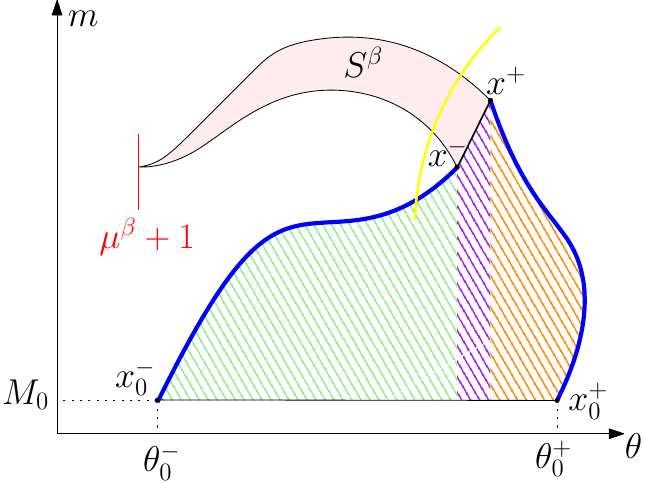}
\end{minipage}
\caption{(Left) Depiction on how to define the integration variables on the intermediate manifold in between two filament pairs. (Right) Geometry of the variables in the main formula, Eq.~\eqref{eq_app:SMV}; thick blue lines represent trajectories of the map connecting $x_0^\pm$ to $x^\pm$, which we depict as continuous for visualization ease. We employ the fact that the map \eqref{eq_app:kicked_rotor_classical} is symplectic in order to calculate $S^\beta$ (in pink) as the sum of green, purple and red shaded areas, which can be easily obtained without the need of complicated line integrals. We also use symplecticity to obtain the Maslov index $\mu^\beta$, not by following $\mathcal{L}_k$, but by the difference between the Maslov indices of the blue trajectories.}
\label{fig_app:geometry}
\end{figure}

The semiclassical mean-value (SMV) formula is given by
\begin{equation}
    \la \widehat{O} \ra_\text{SMV} = \la \widehat{O} \ra_\text{CL}  + 2 \sum_{\beta} \int \text{d} \eta \, \mathcal{W} \big[ \widehat{O} \big](\eta) a_+^\beta( \eta ) a_-^\beta( \eta ) \cos \left( \frac{i}{\hbar} S^\beta ( \eta ) + \frac{i \mu^\beta \pi}{2}  \right) \, ,
    \label{eq_app:SMV}
\end{equation}
where the $\la \widehat{O} \ra_\text{CL}$ term was described in the earlier section.  We shall now provide a geometrical explanation of this formula.

For each filament pair $\beta$  at time $k$, we denote by $\mathcal{L}_k^+$ and $\mathcal{L}_k^-$ the upper and lower filament, and by $\mathcal{L}_k^{\eta}$ a smooth curve lying inbetween the two filaments (see Fig.~\ref{fig_app:geometry}).  The parameter $\eta$ with respect to which integration is performed in Eq.~\eqref{eq_app:SMV} can be any curvilinear coordinate on $\mathcal{L}_k^{\eta}$.
This is illustrated in the left panel of Fig.~\ref{fig_app:geometry}, where the arrows indicate that integration is performed from left to right. 
Note that the precise way in which $\mathcal{L}_k^{\eta}$ is formed is irrelevant in the semiclassical limit, the only requirement being that it lies between $\mathcal{L}_k^+$ and $\mathcal{L}_k^-$ \cite{mittal2020semiclassical}. Each filament ends at the effective edge of our detector, which we take as a circle of radius  $4 \sigma$ centered at $(m_c,\theta_c)$. Numerically each filament is defined by a list of $K$ evenly spaced points $x^+_\kappa \in \mathcal{L}_k^+$ and $x^-_\kappa \in \mathcal{L}_k^-$  ($\kappa = 1,\cdots,K$); from these we form the intermediate manifold $\mathcal{L}_k^{\eta} = \{ (x^+_\kappa + x^-_\kappa)/2 \, ; \, \kappa= 1, \dots, K \}$.

We now move on to describe how to calculate $a_\pm^\beta(\eta)$, $S^\beta(\eta)$ and $\mu^\beta$. Consider the point $x^\eta \in \mathcal{L}_k^{\eta}$, of coordinate $\eta$, and the associated  pair $(x^+(\eta),x^-(\eta))$ on $\mathcal{L}_k^+ \times \mathcal{L}_k^-$ of which  $x^\eta$ is  the center.
These points are the image at time $k$ of a certain pair $(x^+_0,x^-_0)$ of points on the manifold $\mathcal{L}_0$, whose coordinates are $(M_0,\theta^\pm_0)$. The action $S^\beta$ is the area of the curve formed by the path from $x^+$ to $x^-$ on $\mathcal{L}_k$ (which may stretch far outside the detector) and the  line segment joining $x^+$ and $x^-$ (\emph{i.e.}~the chord $x^+ - x^-$), as shown in right panel of Fig.~\ref{fig_app:geometry}. In order to save us the trouble of performing line integrals over $\mathcal{L}_k$, which can be an extremely long manifold, we note that the transformation
\begin{equation}
    (M_0, \theta_0^\pm) = x_0^\pm \longmapsto x^\pm = T^k x^\pm_0
\end{equation}
is a symplectomorphism, and thus the area of the closed curve $\{x_0^- \to x^- \to x^+\to x_0^+ \to x_0^-\}$ build by gluing together the trajectories $x^\pm(t)$ joining $x_0^\pm$ to $x^\pm$, the segment $[x^-,x^+]$ and the segment $[x^+_0,x^-_0]$, is exactly $S^\beta$ \cite{ArnoldBook}. The same idea can be used to calculate the Maslov index $\mu^\beta$: This index counts the number of times the manifold $\mathcal{L}_k$ has a vertical tangent (see Fig.~\ref{fig_app:geometry}), but since it is a topological invariant and the line segments joining $x_0^- \mapsto x_0^+$ and $x^- \mapsto x^+$ do not contribute, we can decompose $\mu^\beta = \mu^+ - \mu^-$ and use the trajectories to obtain it instead of following $\mathcal{L}_k$.

The last piece missing are the coefficients $a_\pm$ in \eqref{eq_app:SMV}. Due to our parametrization of $\mathcal{L}_k^\pm$ in terms of the initial angles, however, the formulas in \cite{mittal2020semiclassical} vastly simplify and the whole integration measure and its Jacobian become
\begin{equation}
    \int \text{d}\eta \, a_+(\eta) \, a_-(\eta) \quad \longrightarrow \quad \frac{1}{2\pi}\int \text{d}\eta \, \sqrt{ \left\vert \frac{\text{d}\theta_0^+}{\text{d} \eta} \frac{\text{d}\theta_0^-}{\text{d} \eta}  \right\vert } \, .
\end{equation}

\section{Herman-Kluk calculations}\label{sec:hk}

\subsection{Herman-Kluk propagator on the circle}

The Herman-Kluk (HK) propagator is expressed in terms of a coherent state (CS) kernel. In order to implement it, one must first be able to write an overcomplete CS basis for the underlying Hilbert space. For the typical $L^2(\mathbb{C})$, in which HK calculations are usually performed, position and momentum expressions for the CS are essentially identical. The kicked rotor system (KRS), however, requires us to formulate quantum mechanics on $L^2(S^1)$, where $S^1 \sim [0, 2\pi)$. In this case, momentum and position are substituted by action (or angular momentum) and angle variables, and constructing the corresponding representations is known to be non-trivial \cite{levy1976afraid}. Since the ``angle operator" on $S^1$ is bounded and continuous, the action ends up unbounded and quantized in multiples of $\hbar$ \cite{GazeauBook}. It is numerically much simpler to work with CSs written in terms of unbounded, quantized variables than continuous bounded ones, which is why we choose to propagate HK wavefunctions in momentum representation \cite{Maitra2000}. The HK wavefunction then reads
%
% \begin{equation}
%     \psi_k(m') = \frac{1}{2\pi \hbar} \int \text{d}m_0 \, \text{d}\theta_0 \, \sqrt{R_{\gamma}(m_k, \theta_k)} \exp \l[ \frac{i}{\hbar}  S(m_k, \theta_k) \r] g_{\gamma}(m', m_k, \theta_k) \mathcal{V}_\gamma(m_0, \theta_0) \, , \label{eq_app:HK}
% \end{equation}
%
\begin{equation}
    \psi_k(m') = \frac{1}{2\pi} \sum_{m_0 \in \hbar \mathbb{Z}}  \int \text{d}\theta_0 \, \sqrt{R_{\gamma}(m_k, \theta_k)} \exp \l[ \frac{i}{\hbar}  S(m_k, \theta_k) \r] g_{\gamma}(m', m_k, \theta_k) \mathcal{V}_\gamma(m_0, \theta_0) \, , \label{eq_app:HK}
\end{equation}
where  $(m_k, \theta_k)^t = T^k (M_0 , \theta_{0})^t$ and, generally for $L^2(S^1)$,
\begin{align}
    R_\gamma(m_k, \theta_k) &= \frac{1}{2} \l[ \frac{\partial m_k}{\partial m_0} + \frac{\partial \theta_k}{\partial \theta_0} - \frac{1}{i \gamma} \l( \frac{\partial m_k}{\partial \theta_0} \r) - i \gamma \l( \frac{\partial \theta_k}{\partial m_0} \r) \r] \\[8pt]
    g_\gamma(m', m_k, \theta_k) &= \l( \frac{\hbar}{\pi \gamma } \r)^\frac{1}{4} \exp \l[ -\frac{1}{2\hbar\gamma} (m' - m_k)^2 - \frac{i}{\hbar} m' \theta_k \r] \\[8pt]
    S(m_k, \theta_k) &= \alpha \sum_{i=1}^k \l[ K(m_{i}) -  V(\theta_{i-1}) \r] \\
    \mathcal{V}_\gamma(m_0, \theta_0) &= 
%    \int_{S^1} \text{d}\theta \, \psi_0(\theta) \, \overline{g_\gamma(\theta,\theta_0,m_0)}  
    \sum_{m \in \hbar \mathbb{Z}} \psi_0(m) \, \overline{g_\gamma(m,m_0,\theta_0)} 
    \, .
\end{align}
For all of our calculations, we chose the unbiased CS width, namely $\gamma=1$, in order to represent a situation in which nothing is known about the system's properties. The variables $(m_k, \theta_k)$ are, as in the main text, obtained by iterating $(m_0, \theta_0)$ using \eqref{eq_app:kicked_rotor_classical}. For the KRS defined in the latter equation, we have $K(m) = m^2/2$ and $V(\theta) = \sin \theta $, and given that our initial state is given by $\ket{M_0}$, the overlap term simplifies to 
\begin{equation}
    \mathcal{V}_\gamma(m_0, \theta_0) = \overline{g_\gamma(M_0, m_0, \theta_0)} \, .
\end{equation}
It is also important to mention that we need to track the branch changes in the pre-factor $\sqrt{R_\gamma(m_k, \theta_k)}$, for which the recipe is very simple:
\begin{equation}
    \text{\, \textbf{IF} \,} \Re \l[ R_\gamma(m_k, \theta_k) \r] < 0 \text{\, \textbf{AND} \,} \Im \l[ R_\gamma(m_k, \theta_k) \r] \times \Im \l[ R_\gamma(m_{k-1}, \theta_{k+1}) \r] < 0 \text{\, \textbf{THEN} \,} \mu = \mu + 1 \, ,
\end{equation}
where $\mu$ is the number of branch crossings. 

\subsection{Sampling the initial trajectories}

Integral \eqref{eq_app:HK} is best performed by Monte Carlo. Importance sampling of the initial trajectories is extremely important in order to avoid wasting computational resources. Sampling according to the overlap $\mathcal{V}_\gamma(m_0, \theta_0)$ is usually a good choice, since it is the only term in \eqref{eq_app:HK} depending on our initial state and trajectories falling far from it will not contribute to the final wavefunction. We then pick the real part of the overlap as the probability density function (PDF), which in essence provides a rule for sampling the initial $m_0$'s only. The $\theta_0$'s need to be sampled uniformly on $[0,2\pi)$ no matter what, such that the final phase-space PDF is given by
\begin{equation}
    \mathbb{P}_0(m_0, \theta_0) = \mathcal{U}_{[0,2\pi)}(\theta_0) \times \exp \l[ - \frac{\l(x - m_0 \r)^2}{2 \gamma \hbar} \r]  \, ,
\end{equation}
where $\mathcal{U}_A$ is the uniform distribution on interval $A$. Note that $\mathbb{P}$ is not normalized, since we will always renormalize the final wavefuction in order to increase the accuracy of the HK method \cite{Maitra2000}. With this choice, \eqref{eq_app:HK} now becomes
\begin{equation}
    \psi_k(m') = \frac{1}{2\pi \hbar} \int \text{d} \l[ \mathbb{P}_0(m_0, \theta_0) \r] \, R_{\gamma}(m_k, \theta_k) \exp \l[ \frac{i}{\hbar}  S(m_k, \theta_k) \r] g_{\gamma}(m', m_k, \theta_k) \exp \l[ \frac{i}{\hbar}  m' \theta_0 \r] \, .
\end{equation}
To compute the corresponding mean values we then follow the exact same procedure as described in Subsec.~\ref{subsec:mean}.

\subsection{Convergence checks}

\begin{figure}
    \centering
    \includegraphics[width=\linewidth]{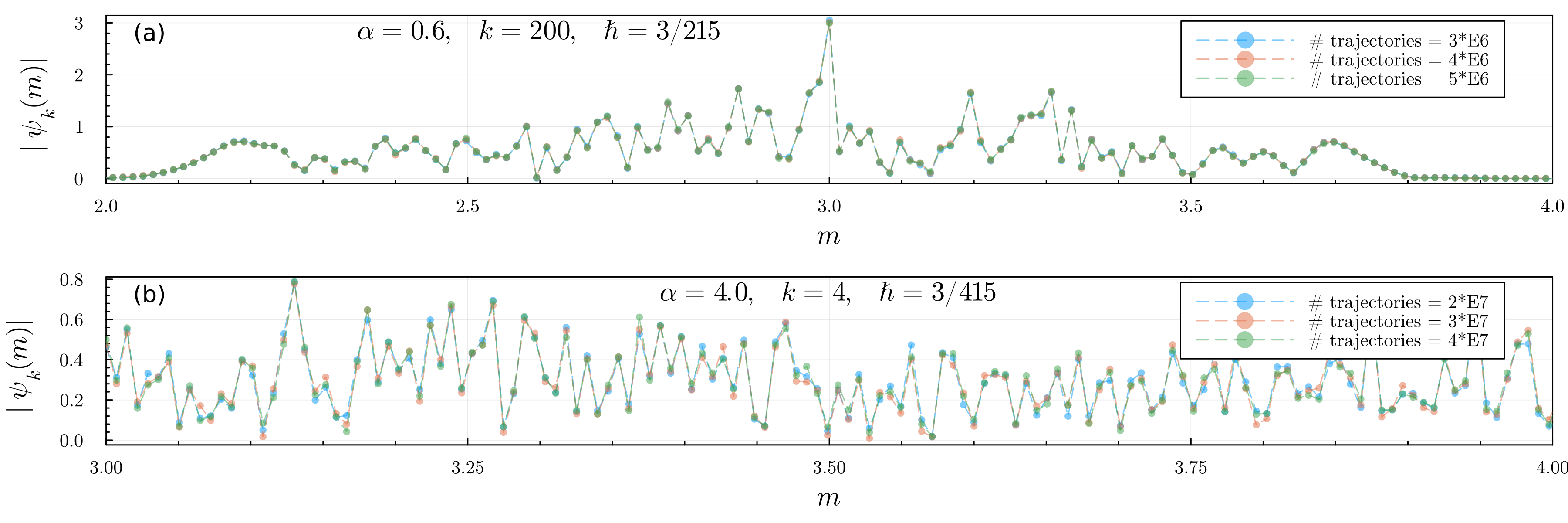}
    \caption{Convergence tests for the HK wavefunction in angular momentum space. Panels (a) and (b) refer to the near-integrable and deep chaotic regimes, respectively; to ease visualization of these highly oscillatory wavefunctions we display only a small interval in angular momentum. Note that achieving a decent convergence on the deep chaotic regime requires 10 times more trajectories than in the near-integrable one.}
    \label{fig_app:wv_convergence}
\end{figure}

\begin{figure}
    \centering
    \includegraphics[width=\linewidth]{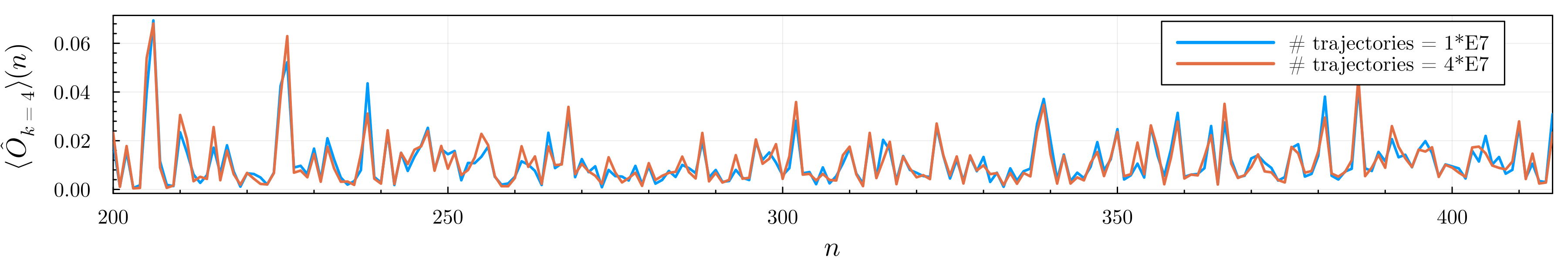}
    \caption{Mean values in the deep chaotic regime for two different values of the number of trajectories used in the HK integral, as a function of time. The parameters are the same as in Fig.~\ref{fig_app:wv_convergence}(b), with the same detector as in the main text. Qualitative behavior of the mean values in unchanged despite using four times more trajectories, showing that the results presented in the main text are properly converged.}
    \label{fig_app:meanvals_convergence}
\end{figure}

Achieving convergence in HK results becomes increasingly hard as the amount of chaos in the system grows. For near-integrable systems, however, convergence is relatively unexpensive. In Fig.~\eqref{fig_app:wv_convergence} we display convergence checks for the HK wavefunction in momentum space using an increasing number of trajectories. We provide checks for both the near-integrable and the deep chaotic cases used in the paper, choosing the smallest $\hbar$ used for either case. This check is enough because initial value representations such as the HK propagator rely on phase cancellation between non-stationary trajectories, and such cancellations are very sensitive to $\hbar$: The smallest its value, the more oscillatory the exponentials that need to cancel up inside integral \eqref{eq_app:HK} (which, numerically, becomes a sum over the contribution of every individual trajectory). 

It can be seen in Fig.~\ref{fig_app:wv_convergence}(b) that convergence was achieved with small deviations, which although larger than the ones seen on Fig.~\ref{fig_app:wv_convergence}(a), are significantly smaller than the size of the signal itself. Mean values computed from the wavefunctions in Fig.~\ref{fig_app:wv_convergence}(b) will not differ from each other as significantly as needed in order to correct for the noise-like data presented in the main text. Indeed, as can be seen in Fig.~\ref{fig_app:meanvals_convergence}, mean values in the deep chaotic regime calculated with $1 \times 10^7$ and $4 \times 10^7$ show that the latter are only marginally  different from the former, and that the qualitative behavior of the mean values is unaltered.

\section{Semiclassical mean values on the torus for mixed and chaotic dynamics}

The method described in the main text and in Sec.~\ref{sec:smv} is by no means restricted to the kicked rotor. In this section we provide a further employment of our method to the paradigmatic kicked Harper model, which being defined on the torus has both its ``position'' and ``momentum'' discretized in units of $\hbar$. Due to intrinsic difficulties in dealing with coherent-state representations on the torus \cite{GazeauBook} the HK propagator was never adapted to this setting, such that we do not include comparisons with it in this section. 

Kicked Harper model is taken here as
\begin{equation}
\begin{cases}
    \varphi_k = \varphi_{k-1} + \alpha \, \sin \theta_{k-1} \quad \text{mod}_\pm \, \pi \\
    \,\, \theta_k = \theta_{k-1} - \alpha \, \sin \varphi_k \qquad \text{mod}_\pm \, \pi  \, .
\end{cases} 
    \label{eq:harper_map_classical}
\end{equation}
We take $\theta_0 \in [-\pi, \pi)$ and $\varphi_0 \in [-\pi, \pi)$, with the modulo defined as
\begin{equation}
    x \, \text{mod}_\pm \, \pi = 
    \begin{cases}
        x + 2\pi \, , \quad x < -\pi \\
        x - 2\pi \, , \quad x \geq \pi 
    \end{cases} \, .
\end{equation}
Thus, $(\varphi_k, \theta_k)$ are always contained in the flat torus $[-\pi, \pi) \times [-\pi, \pi)$, the above modulo being responsible for identifying its boundaries. An important difference between the kicked rotor and the kicked Harper model is that, in the former, the Maslov index of the propagated Lagrangian manifold always increases by one when crossing a caustic (see Fig.~\ref{fig_app:geometry}), while in the latter the orientation of the caustic crossing needs to be accounted for. Clockwise and counter-clockwise crossings increase or decrease the Maslov index by one, respectively \cite{creagh1990geometrical}. 

The quantum analogue to \eqref{eq:harper_map_classical} is exactly given by the map
\begin{equation}
    \vert \psi_k \rangle =  \exp \left( \frac{i \, \alpha \, \widehat{\cos \varphi}}{\hbar} \right) \exp \left( \frac{i \, \alpha \, \widehat{\cos \theta}}{\hbar} \right) \vert \psi_{k-1} \rangle \, ,
    \label{eq:harper_map_quantum}
\end{equation}
where quantum propagation follows exactly the same strategy as presented in Sec.~\ref{sec:qu}, with integrals exchanged by sums when necessary. Note that the general formula for the Wigner function on the torus is also slightly modified with respect to the one on the cylinder \cite{rivas1999weyl, miquel2002quantum}. Double periodicity in the quantum domain now forces $\hbar$ to be quantized as $\hbar = 2\pi/N$, where $N$ is an integer. Since now both $\theta$ and $\varphi$ representations are quantized in units of $\hbar$, entering the semiclassical regime is tantamount to increasing ``resolution'' (that is, $N$) in either of these representations. We evolve the same initial state as in the main text, but now assume the initial state has $M_0 \equiv \Phi_0 = \pi/2$. Enforcing this condition means that we need to adapt $N$ as
\begin{equation}
    \Phi_0 = \frac{\pi}{2} = n_0 \hbar = \frac{2 \pi n_0}{N} \quad \Longrightarrow \quad N = 4 n_0 \, ,
\end{equation}
where $n_0$ represents how many multiples of $\hbar$ emerge from the quantization of the initial manifold's ``momentum'', $\Phi_0$. With these definitions we can keep the classical dynamics fixed as we vary $N$, which is always advantageous for semiclassical investigations and was also pursued in our treatment of the kicked rotor (see main text).

\begin{figure}
    \centering
    \includegraphics[width=\linewidth]{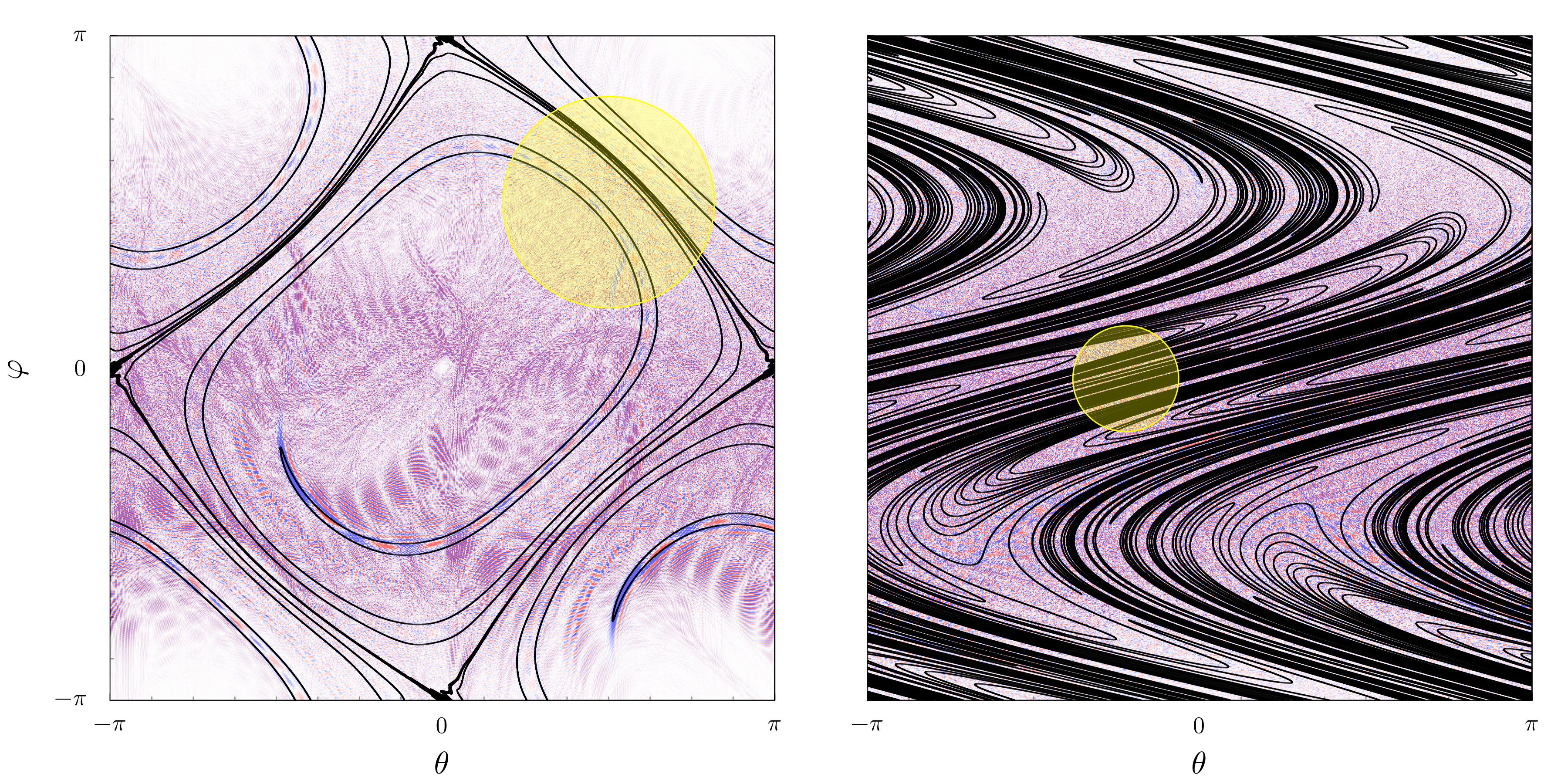}
    \caption{Initial Lagrangian manifold $\Phi_0 = \pi/2$ and its associatetd wave function, $\varphi(\theta) = \exp(i \Phi_0 \theta/\hbar)/\sqrt{2\pi}$, $\hbar=2\pi/N$, $N=800$, is evolved on the classical torus and on its quantized equivalent. Quantum datum is here visualized by means of Wigner functions, in the same spirit as Fig.~1 in the main text, allowing for a direct comparison with the manifold's evolution in phase space. Left panel uses $\alpha=1.0$ and $k=25$ in \eqref{eq:harper_map_classical}, while the right panel has $\alpha=4.0$ and $k=3$. The detectors over which mean values will be taken can be seen in ỳellow, having different radii but being both truncated at 4 standard deviations.}
    \label{fig:harper_wig_vs_lag}
\end{figure}

Fig.~\ref{fig:harper_wig_vs_lag} provides examples of the kicked Harper model in two different dynamical regimes, comparing the evolved Lagrangian manifold $\varphi(\theta)=\pi/2$ with is corresponding Wigner-function analogue in phase space. The figure also contains detectors at specific locations, following the same line of reasoning as Fig.~1 in the main text. In the left panel a homoclinic figure is clearly visible connecting the zeros in each boundary, while the spiraling whorl around the origin denounces the presence of a stable fixed point at $(\varphi, \theta) = (0,0)$. Simultaneous presence of a homoclinic orbit and stable fixed points essentially defines the \emph{mixed} regime, in which chaotic and near-integrable dynamics coexist. Right panel shows a typical case of strong chaotic behavior in which no stable fixed points are visible.

\begin{figure}
    \centering
    \includegraphics[width=\linewidth]{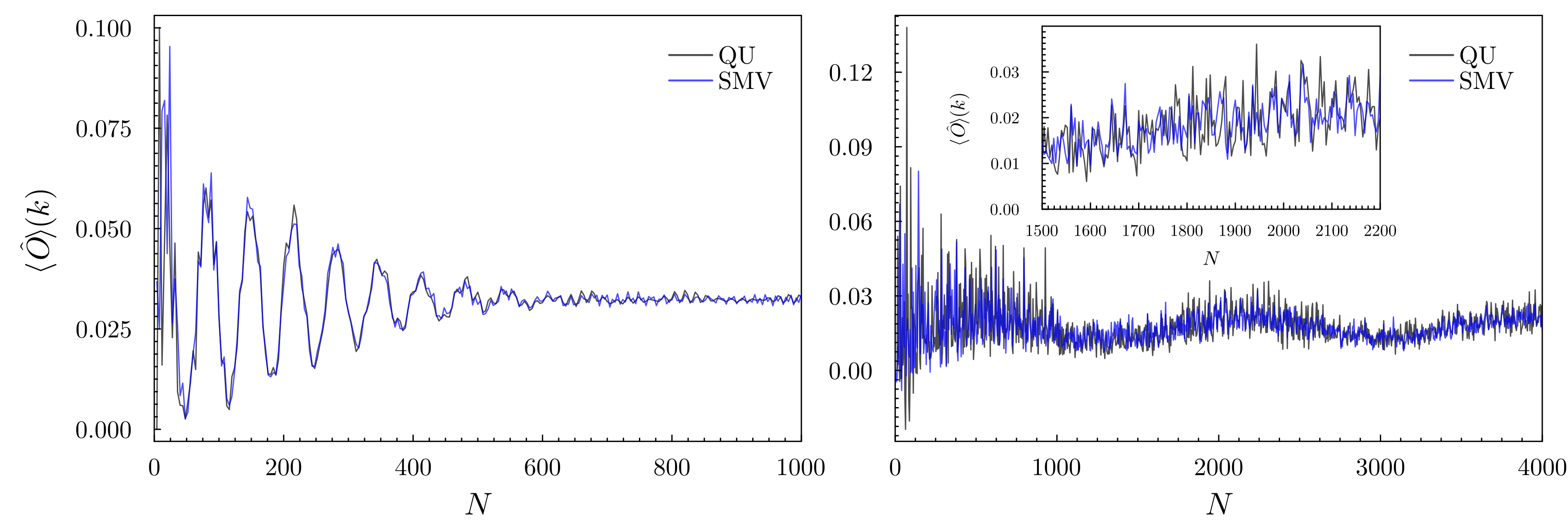}
    \caption{Mean values associated with the two panels in Fig.~\ref{fig:harper_wig_vs_lag} as a function of $\hbar = 2\pi/N$. A zoom on the right panel is provided in the inset.}
    \label{fig:mean_vals_harper}
\end{figure}

As discussed in the main paper, our current implementation of the semiclassical mean value method is optimized for dealing with filament sections that are approximately straight lines, while in Fig.~\ref{fig:harper_wig_vs_lag} the sections are clearly curved. A more serious problem is the fact that there are 6 small ``fingers'' within the detector, \emph{i.e.}~filament sections that come out from the same side as the come in and require a more delicate treatment in terms of Airy functions. Despite these complications, the mean values computed from our method come out extremely accurate, as can be seen in the left panel of Fig.~\ref{fig:mean_vals_harper}. The deep chaotic regime in the right panel of Fig.~\ref{fig:harper_wig_vs_lag} is extremely hard to deal with but our method is still remarkably accurate, as can be seen in the right panel of the aforementioned figure.

\section{Code companion}

Our code is made public and is available at \url{https://github.com/gabrielmlando/MeanVals}. It is written fully in Julia, and we include a minimal working example for each module used for producing the data in the main text in the form of a Jupyter Notebook. We do not provide the original data due to size limitations. We encourage anyone interested in further optimizing/generalizing our code to do so, as long as credit is given by citing the original paper.

\bibliography{bib.bib}